\documentclass[12pt]{article}
\usepackage{epsfig}

\usepackage{a4}

\usepackage{amsmath}

\begin{document}
\title{
Chern-Simons Gravities (CSG) and Gravitational Chern-Simons (GCS) densities in all dimensions
}
\author{{\large }
{\large D. H. Tchrakian}$^{\star\dagger\ddagger}$ \\ \\
$^{\dagger}${\small School of Theoretical Physics, Dublin Institute for Advanced Studies,
10 Burlington Road, Dublin 4, Ireland}\\ 
$^{\dagger}${\small Department of Computer Science, Maynooth University, Maynooth, Ireland}\\
$^{\ddagger}${\small Theory Division, Yerevan Physics Institute, Alikhanian Br. Str. 2, Yerevan, Armenia}}

\date{}
\newcommand{\dd}{\mbox{d}}
\newcommand{\tr}{\mbox{tr}}
\newcommand{\la}{\lambda}
\newcommand{\bt}{\beta}
\newcommand{\del}{\delta}
\newcommand{\ep}{\epsilon}
\newcommand{\ta}{\theta}
\newcommand{\ka}{\kappa}
\newcommand{\f}{\phi}
\newcommand{\vf}{\varphi}
\newcommand{\F}{\Phi}
\newcommand{\al}{\alpha}
\newcommand{\ga}{\gamma}
\newcommand{\de}{\delta}
\newcommand{\si}{\sigma}
\newcommand{\Si}{\Sigma}
\newcommand{\bnabla}{\mbox{\boldmath $\nabla$}}
\newcommand{\bomega}{\mbox{\boldmath $\omega$}}
\newcommand{\bOmega}{\mbox{\boldmath $\Omega$}}
\newcommand{\bsi}{\mbox{\boldmath $\sigma$}}
\newcommand{\bchi}{\mbox{\boldmath $\chi$}}
\newcommand{\bal}{\mbox{\boldmath $\alpha$}}
\newcommand{\bpsi}{\mbox{\boldmath $\psi$}}
\newcommand{\brho}{\mbox{\boldmath $\varrho$}}
\newcommand{\beps}{\mbox{\boldmath $\varepsilon$}}
\newcommand{\bxi}{\mbox{\boldmath $\xi$}}
\newcommand{\bbeta}{\mbox{\boldmath $\beta$}}
\newcommand{\ee}{\end{equation}}
\newcommand{\eea}{\end{eqnarray}}
\newcommand{\be}{\begin{equation}}
\newcommand{\bea}{\begin{eqnarray}}

\newcommand{\ii}{\mbox{i}}
\newcommand{\e}{\mbox{e}}
\newcommand{\pa}{\partial}
\newcommand{\Om}{\Omega}
\newcommand{\om}{\omega}
\newcommand{\vep}{\varepsilon}
\newcommand{\bfph}{{\bf \phi}}
\newcommand{\lm}{\lambda}
\def\theequation{\arabic{equation}}
\renewcommand{\thefootnote}{\fnsymbol{footnote}}
\newcommand{\re}[1]{(\ref{#1})}
\newcommand{\R}{{\rm I \hspace{-0.52ex} R}}
\newcommand{\N}{{\sf N\hspace*{-1.0ex}\rule{0.15ex}%
{1.3ex}\hspace*{1.0ex}}}
\newcommand{\Q}{{\sf Q\hspace*{-1.1ex}\rule{0.15ex}%
{1.5ex}\hspace*{1.1ex}}}
\newcommand{\C}{{\sf C\hspace*{-0.9ex}\rule{0.15ex}%
{1.3ex}\hspace*{0.9ex}}}
\newcommand{\eins}{1\hspace{-0.56ex}{\rm I}}
\renewcommand{\thefootnote}{\arabic{footnote}}

\maketitle


\bigskip

\begin{abstract}
Chern-Simons gravities and gravitational Chern-Simons densities are constructed using the non-Abelian Yang-Mills
Chern-Simons densities. As such, they are defined only in odd dimensions. We propose instead an analogous construction
employing what we term Higgs--Chern-Simons (HCS) densities,
which are defined in $all\ dimensions$. This enables the definition of extended versions of Chern-Simons gravities in all dimensions.
Employing the same prescription, the definition of gravitational Chern-Simons densities is extended to all even dimensions, but only to
$4p-1$ $odd$ dimensions. All our considerations are restricted to $vacuum$ fields.
\end{abstract}

\bigskip
\bigskip
\bigskip
\noindent
Talk given at XVII International Conference on Symmetry Methods in Physics (Symphys XVII), Yerevan, Armenia, 09-15 July, 2017.

\section{Introduction}
In the context of this report, Chern-Simons gravities (CSG) and gravitational Chern-Simons (GCS) densities are distinct
but not unrelated objects. They both result from the definition of Chern-Simons (CS) densities of the non-Abelian (nA) Yang-Mills (YM) fields.
The CSG consist of some superpositioxn of $usual$ gravitational Lagrangians displaying all orders of the Riemann curvature, including the $0$-th
- the cosmological constant. The GCS are the direct gravitational analogues of the nA CS densities, and like these, they can be employed
in conjuction with $usual$ gravitational Lagrangians.

CS gravity (CSG) in $2+1$ dimensions was proposed in
\cite{Witten:1988hc} and was subsequently extended to $2n+1$ dimensions in \cite{Chamseddine:1989nu,Chamseddine:1990gk}.
Gravitational CS densities in $2+1$ dimensions first appeared~\footnote{In these references the CS density is used to
generate mass in a (Topological) non-Abelian field theory, and in the gravitational case the GCS density is added to the
usual gravity.} in~\cite{Deser:1982vy,Deser:1981wh} and were extended to
odd dimensions in \cite{Mardones:1990qc,Zanelli:2012px}.

Since the construction of both CSG's and GCS densities employ the nA CS densities, it follows that these are defined in
odd dimensions only. The main aim here to extend these definitions to cover also even dimensions. Indeed,
such examples are present in the literature both for CSG~\cite{Chamseddine:1990gk,MacDowell:1977jt}, and,
for GCS density~\cite{Jackiw:2003pm,Araneda:2016iiy} in $3+1$ dimensions.

For CSG systems in $3+1$ dimensions, two such proposals are made, in
\cite{MacDowell:1977jt} and in \cite{Chamseddine:1990gk}. In the second of these, in~\cite{Chamseddine:1990gk}, a
Higgs-like~\footnote{The scalar used in \cite{Chamseddine:1990gk}, and in the rest of this presentation, is not a matter field.
We refer to it as Higgs because of its provenance $via$ the dimensional descent of a non-Abelian (nA) density. The gravitational
degrees of freedom it yields is technically related to the nA Higgs field as the spin-connection is related to the nA curvature.} scalar
is employed, which is contained in what is proposed in the present report. A GCS densities in $3+1$ dimensions are introduced in
\cite{Jackiw:2003pm} and \cite{Araneda:2016iiy}, which are employed for the purpose of modifying the usual (Einstein) gravity.
In \cite{Jackiw:2003pm}, the (gravitational) Pontryagin density is employed, while in \cite{Araneda:2016iiy} both the Pontryagin and
the Euler densities are employed. Definitions of GCS densities
are made in this spirit in what follows here.

The pivotal point in our considerations is the explotation of (what is referred to here) Higgs--Chern-Simons (HCS) densities,
which are defined in all, odd and even dimensions. These result from the dimensional descent of Chern-Pontryagin (CP) densities from
some even dimension down to any even or odd dimension. They are introduced in \cite{Tchrakian:2010ar,Radu:2011zy}
and in Appendix A of \cite{Tchrakian:2015pka}. These HCS densities will be employed instead of the usual CS densities, to
construct gravitational models in all dimensions which might be described as Higgs--Chern-Simons gravities (HCSG), and also to
construct gravitational Higgs--Chern-Simons densities (GHCS) in all even dimensions, and in $4p-1$ odd dimensions.
It is known that GCS densities are absent~\cite{Mardones:1990qc,Zanelli:2012px} in $4p-3$ dimensions, and it will turn out that
also GHCS densities are absent in $4p-1$ dimensions.

All statements made are illustrated $via$ typical exmples, and
all calculations are carried out in the Einstein-Cartan formulation of gravity.
In the Einstein-Cartan formulation, the gravitational system can be described in terms
of ({\bf a}) the $Vielbein$ fields $e_M^a$ (and its inverse $e^M_a$) which are related to the metric through
\[
g_{MN}=e_M^ae_N^b\,\eta_{ab}\ ,\quad g^{MN}=e^M_ae^N_b\,\eta^{ab}\,,
\]
$\eta_{ab}$ being the flat (Minkowskian or Euclidean) metric, and ({\bf b})
the spin-connection $\om_M^{ab}$ which defines the Riemann curvature
\[
R_{MN}=\pa_{[M}\om_{N]}^{ab}+(\om_{[M}\om_{N]})^{ab}\,.
\]
Since the frame indices $a,b,\dots$ are raised/lowered by a flat metric, and since we are not particularly concerned with the signature
of the spaces(s), we will henceforth not pay any attention to whether the frame index is covariant or contravariant.

It is the identification of the spin-connection
and the Riemann curvature with the YM connection and curvature in $D$ dimensions, through
\be
\label{con1}
A_{M}=-\frac12\,\om_{M}^{ab}\,L_{ab} 
\Rightarrow F_{MN}=-\frac12\,R_{MN}^{ab}\,L_{ab}\,,\quad M=1,2,\dots,D\ ,\quad a=1,2,\dots,D
\ee
that is exploited in the definitions of both GCS densities and CSG gravities.
The matrices $L_{ab}$ in \re{con1} are representations of $SO(D)$.
In what follows, we will employ the Dirac (Clifford algebraic) representations for $L_{ab}$
\be
\label{L}
L_{ab}=\ga_{ab}=-\frac14\,[\ga_a,\ga_b]\,,
\ee
in terms of $\ga_{a}$, the gamma matrices in $D$ dimensions.

The Chern-Simons (CS) density expressed in terms of the YM connection $A_M$ and curvature $F_{MN}$
is defined through the one-step descent of the Chern-Pontryagin (CP) density
\[
\Om_{\rm CP}= \mbox{Tr}\, F\wedge F\wedge \dots\wedge F=\bnabla\cdot\bOmega\ ,\quad 2n\ {\rm times}
\]
in some $even$, $D=2n$ dimensions~\cite{Jackiw:1985}.
The one-step descent in question is a result of the fact that the CP density $\Om_{\rm CP}=\bnabla\cdot\bOmega$ is
a total divergence. The CS density is defined
as any one component of the vector-valued density $\bOmega$, say $\Omega_D$, $i.e.$ $\Omega_{\rm CS}=\Omega_D$, which
now depends only on the $D-1$ coordinates $x_\mu\ ,\quad\mu=1,2,\dots,D-1$. Hence, the CS density thus defined necessarily
exists in some $odd$ dimension  $2n-1$. The CS densities are by construction $gauge\ variant$ but their variational equations turn
out to be $gauge\ covariant$, and display other interesting features
prominent among which is their gauge transformation properties in the case of nA fields,
leading to their exploitation in quantum field theory~\cite{Deser:1982vy,Deser:1981wh}.
These aspects of CS theory will not be pursued here. Instead,
our aim here is to exploit the non-Abelian (nA) CS densities to construct gravitational-CS (GCS) densities and
Chern-Simons gravities (CSG).

Already at this stage it is clear that the passage from YM to gravity prescribed by \re{con1} is problematic in the context of
gravitational CS (GCS) densities and CS gravities (CSG), since these are defined
in $D-1$ dimensions with coordinates $x_\mu\ ,\quad\mu=1,2,\dots,D-1$, while the frame indices run over
$a=1,2,\dots,D$ instead of $D-1$. This discrepancy is corrected in each case respectively, GCS and CSG,
by sharpening the prescription \re{con1}.

Following this prescription, it is clear that GCS densities and CSG systems can be defined only in {\bf odd} dimensions.
Our aim in this presentation is to propose GCS densities and CSG systems in all dimensions, namely in both {\bf odd}
and {\bf even} dimensions. To this end, one starts from a version of the Chern-Pontryagin (CP) densities that is defined in odd
and even dimensions. These are the Higgs--CP (HCP) densities resulting from the dimensional descent of a CP density in some even
dimension, down to some odd or even residual dimension. The Higgs field in this case is a relic of the YM connection in the higher
dimension and these HCP densities are $total\ divergences$ like the CP densities.

The dimensional reduction of gauge fields has a long history~\cite{Forgacs:1979zs,Schwarz:1981mb,Kapetanakis:1992hf}. The calculus of dimensional reduction used
in \cite{Tchrakian:2010ar,Radu:2011zy,Tchrakian:2015pka} is
an extended version of that of Ref.~\cite{Schwarz:1981mb}. In most applications of this calculus, the descent was carried
out on the Yang-Mills action/energy density in higher dimensons. Application to the descent of Chern-Pontryagin densities was first carried out in~\cite{Sherry:1984ky}
applied to the third and fourth  CP denisties in $6$ and $8$ dimensions down to $3$ dimensions, yielding the monopole (topological) charge densities of two extended
Yang-Mills--Higgs (YMH) theory on $\R^3$. Soon after in \cite{OBrien:1988whk},
the fourth CP denisty in $8$ dimensions was dimensionally reduced down to $4$ dimensions, yielding the monopole~\footnote{In \cite{OBrien:1988whk}, the descent was performed
both on the $4$-th CP density and the $p=2$ YM~\cite{Tchrakian:1984gq} system to yield a YMH theory supporting ``instantonss'' on $\R^4$.}  (topological) charge density of a YMH theory on $\R^4$.
Subsequently this formulation was extended to all even and odd dimensions in Refs.~\cite{Tchrakian:2010ar,Radu:2011zy}
and in Appendix A of Ref~\cite{Tchrakian:2015pka}.

The $total\ divergence$ property of the HCP densities enables, $via$ the standard one-step descent, the definition of CS densities in {\bf all} dimensions.
We refer to these as Higgs--CS (HCS) densities resulting from the one-step descent - from $D$ to $D-1$ dimensions. In $3+1$ dimensions in particular,
two such HCS densitied were employed in Refs.~\cite{Navarro-Lerida:2013pua,Navarro-Lerida:2014rwa} in $SO(5)$ and $SU(3)$ YMH models.

In arbitrary dimensions, the HCS densities are introduced in \cite{Tchrakian:2010ar,Radu:2011zy,Tchrakian:2015pka}. Recently, some HCS densities
were given independently in \cite{Szabo:2014zua}. While in many cases the HCS densities of \cite{Szabo:2014zua} agree with ours~\cite{Tchrakian:2010ar,Radu:2011zy,Tchrakian:2015pka},
they differ most markedly in that they are defined in $odd$ dimensions only, while in our case $all$, even and odd, dimensions are included. The reason for this is that
in \cite{Szabo:2014zua} it is the CS density~\footnote{Dimensional reduction being a calculus of symmetry imposition, it is unsafe to carry it out on a
$gauge\ variant$ density. That in some examples this may not be problematic~\cite{Romanov:1978ur}, happens to be true.}
in the (higher) odd dimensions which is subjected to dimensional reduction,
and then by $2$ dimensions or or by some other even dimension. Thus in the case of \cite{Szabo:2014zua}, only even dimensional HCS are defined.  In our case by
contrast, it is the $gauge\ invariant$ CP density that is subjected to dimensional reduction by any number of dimensions, resulting in HCS densities in odd dimensions.
It turns out that in even residual dimensions, the HCS density thus obtained is $gauge\ invariant$ like the CP density in the bulk.

It is the HCS densities that are employed in constructing the gravitational HCS (GHCS) densities and the HCS gravities (HCSG) in
both odd and even dimensions, applying some variants of the prescription \re{con1}. It turns out that the construction of HCSG
systems is unique, in the sense that in {\bf odd} dimensions where CSG systems exist already before the introduction of the Higgs
field, this CSG system is embedded in the corresponding HCSG system. The situation with the construction of the GCSH density is
rather more $ad\ hoc$, such that in {\bf odd} dimensions where both GCS and GHCS densities can be constructed, the two results
are different. We have thus relegated the subject of GCS and GHCS to an Appendix.

The presentation is organised as follows. In Section {\bf 2} the main building blocks, namely the CS densities for $d=3,5,7$
and HCS densities for for $d=3,4,5,6.7$, dimensions $d=D-1$ are listed. In Section {\bf 3} some CS gravities (CSG) and HCS gravities
(HCSG) are presented. In Section {\bf 4} some gravitational  CS desities (GCS) and gravitational HCS (GHCS) densities are presented.
The reason for the chosen order of Sections {\bf 3} and {\bf 4} is that in the context adopted here,
the GCS (and GHCS) densities are objects which should be employed to modify the more fundamental CSG (and HCSG) systems.

\section{Chern-Simons and Higgs--Chern-Simons densities}
In this Section we present the usual Chern-Simons (CS) densities, which are defined in odd dimensions only,
and the Higgs--Chern-Simons densities which are defined in all dimensions. They are defined for arbitrary gauge
group but the choice of gauge group appropriate for the passage of Yang-Mills to gravity, is specified.
\subsection{The $usual$ Chern-Simons densities in odd dimensions}
The $usual$ Chern-Simons (CS) density in $d$ dimensions results from the one-step descent of the the Chern-Pontryagin (CP)
density in $D=d+1$
dimensions, $D=2n$ being even. Since our final aim is to transition from Yang-Mills to gravity, the gauge group of
the non-Abelian field, is fixed by the prescription \re{con1} where $L_{ab}$ takes its values in the algebra of $SO(D)$ ($D=2n$)
with $a=1,2,\dots, 2n$. The corresp[onding gravitational density is
constructed by evaluating the trace in the CS formula.
Since $D$ is even, $L_{ab}$ are represented the Dirac matrices $\ga_a$, where these are augmented by the chiral matrix $\ga_{D+1}$.
One has then the option of including $\ga_{D+1}$ in the trace of the nA CS density.

For the sake of illustration, we state the CS densities in $d=3,5,7$. Including $\ga_{D+1}$ in the trace these are
\bea
\Omega_{\rm CS}^{(3)}&=&\vep^{\la\mu\nu}\mbox{Tr}\,\ga_5
A_{\la}\left[F_{\mu\nu}-\frac23A_{\mu}A_{\nu}\right]\label{CS32}\\
\Omega_{\rm CS}^{(5)}&=&\vep^{\la\mu\nu\rho\si}\mbox{Tr}\,\ga_7
A_{\la}\left[F_{\mu\nu}F_{\rho\si}-F_{\mu\nu}A_{\rho}A_{\si}+
\frac25A_{\mu}A_{\nu}A_{\rho}A_{\si}\right]\label{CS52}
\\
\Omega_{\rm CS}^{(7)}&=&\vep^{\la\mu\nu\rho\si\tau\ka}
\mbox{Tr}\,\ga_9A_{\la}\bigg[F_{\mu\nu}F_{\rho\si}F_{\tau\ka}
-\frac45F_{\mu\nu}F_{\rho\si}A_{\tau}A_{\ka}-\frac25
F_{\mu\nu}A_{\rho}A_{\si}F_{\tau\ka}\nonumber\\
&&\qquad\qquad\qquad\qquad\qquad\qquad
+\frac45F_{\mu\nu}A_{\rho}A_{\si}A_{\tau}A_{\ka}-\frac{8}{35}
A_{\mu}A_{\nu}A_{\rho}A_{\si}A_{\tau}A_{\ka}\bigg]\,,\label{CS72}
\eea
This choice is appropriate for the construction of CS gravities~\cite{Witten:1988hc,Chamseddine:1989nu,Chamseddine:1990gk} (CSG).
It results in Euler type gravitational densities~\cite{Obukhov:1995eq}.

The other choice is to exclude $\ga_{D+1}$ in the trace. The resulting CS densities 
\bea
\hat\Omega_{\rm CS}^{(3)}&=&\vep^{\la\mu\nu}\mbox{Tr}\,
A_{\la}\left[F_{\mu\nu}-\frac23A_{\mu}A_{\nu}\right]\label{CS31}\\
\hat\Omega_{\rm CS}^{(5)}&=&\vep^{\la\mu\nu\rho\si}\mbox{Tr}\,
A_{\la}\left[F_{\mu\nu}F_{\rho\si}-F_{\mu\nu}A_{\rho}A_{\si}+
\frac25A_{\mu}A_{\nu}A_{\rho}A_{\si}\right]\label{CS51}
\\
\hat\Omega_{\rm CS}^{(7)}&=&\vep^{\la\mu\nu\rho\si\tau\ka}
\mbox{Tr}\,A_{\la}\bigg[F_{\mu\nu}F_{\rho\si}F_{\tau\ka}
-\frac45F_{\mu\nu}F_{\rho\si}A_{\tau}A_{\ka}-\frac25
F_{\mu\nu}A_{\rho}A_{\si}F_{\tau\ka}\nonumber\\
&&\qquad\qquad\qquad\qquad\qquad\qquad
+\frac45F_{\mu\nu}A_{\rho}A_{\si}A_{\tau}A_{\ka}-\frac{8}{35}
A_{\mu}A_{\nu}A_{\rho}A_{\si}A_{\tau}A_{\ka}\bigg]\,,\label{CS71}
\eea
which are appropriate for the construction of gravitational CS densities, and result in Pontryagin
type gravitational densities~\cite{Obukhov:1995eq}..

\subsection{The $Higgs$ Chern-Simons densities in all dimensions}
The definitions of the Higgs--Chern-Simons (HCS) densities however are more ubiquitous. Thus for example a HCS density in
$d$ dimensions, which is derived from the Higgs--Chern-Pontryagin (HCP) density in $D=d+1$, $D$ being even {\bf or} odd.
The HCP density in question, itself 
arises from the dimensional descents of the CP in some even dimension $N\ge d+1$.

The HCP density in $D$ density employed, say $\Om_{\rm HCP}^{D,N}$, is descended from a CP density $\Om_{\rm CP}$ in even $N$ dimensions.
It may thus be useful to denote the HCS density gotten $via$ the one-step descent as
\[
\Om_{\rm HCS}^{(d,N)}\ ,\quad d=D-1\,.
\]
A detailed description of HCP and HCS description is given in \cite{Tchrakian:2010ar,Tchrakian:2015pka}.

In the notation used below, the (suare matrix valued) Higgs scalar $\F$ has dimension $L^{-1}$, as does also the constant $\eta$ which
is the inverse of the sphere over which the descent is carried out.

We list two such the HCS densities in $d=3$, arrived at from the one-step descents from the HCP densities in $D=4$,
each of them descended from the CP densities in $6$ and $8$ dimensions respectively,
\bea
\Omega^{(3,6)}_{\rm HCS}&=&
-2\eta^2\Omega_{\rm CS}^{(3)}-\vep^{\mu\nu\la}\mbox{Tr}\,\ga_5D_{\la}\F\left(F_{\mu\nu}\,\F+F_{\mu\nu} \F\right)\,.\label{HCS36}\\
\Omega^{(3,8)}_{\rm HCS}&=&6\eta^4\Omega_{\rm CS}^{(3)}-\vep^{\mu\nu\la}\,\mbox{Tr}\,\ga_5\,\bigg\{
6\,\eta^2\left(\F\,D_{\la}\F-D_{\la}\F\,\F\right)\,F_{\mu\nu}\nonumber\\
&&\hspace{20mm}-\bigg[\left(\F^2\,D_{\la}\F\,\F-\F\,D_{\la}\F\,\F^2\right)-2\left(\F^3\,D_{\la}\F-D_{\la}\F\,\F^3\right)\bigg]F_{\mu\nu}
\bigg\}\,.\label{HCS38}
\eea
Note that the leading term in both \re{HCS36} and \re{HCS38} is the CS density \re{CS32}.

The HCS density in $d=5$ arrived at from the one-step descent of the HCP density in $D=6$,
itself descended from the CP density in $8$ dimensions, is
\bea
\Omega^{(5,8)}_{\rm HCS}&=&2\eta^2\Om_{\rm CS}^{(5)}+\vep^{\mu\nu\rho\si\la}\,\mbox{Tr}\,\ga_7\bigg[
D_{\la}\F\left(\F F_{\mu\nu}F_{\rho\si}+F_{\mu\nu}\F F_{\rho\si}+F_{\mu\nu}F_{\rho\si}\F\right)\bigg]\label{HCS58}
\eea
and the HCS density in $d=7$ arrived at from the one-step descent of the HCP density in $D=8$, itself descended from the CP
density in $10$ dimensions, is
\bea
\Omega^{(7,10)}_{\rm HCS}&=&\eta^2\Om_{\rm CS}^{(7)}
+\vep^{\mu\nu\rho\si\tau\la\ka}\,\mbox{Tr}\,\ga_9D_{\ka}\F\,F_{\mu\nu}F_{\rho\si}(F_{\tau\la}\,\F+\F\,F_{\tau\la})\,.\label{HCS710}
\eea
Note that the leading term in \re{HCS58} is the CS density \re{CS52}, and that in \re{HCS710} the CS density \re{CS72}.

Thus, in all odd dimensions where both a CS and a HCS density exist, the leading term in the HCS density $\Omega^{(d,N)}_{\rm HCS}$
in (odd) $d$ dimensions that pertains to the HCP density desecnded from the CP density in (even) $N$ dimensions, is the
CS density $\Omega_{\rm CS}^{(d)}$.

The situation is as expected, entirely different for HCS densities in even dimensions, where there are no $usual$
CS densities. In those cases the HCS densities are expressed {\bf entirely} in terms of the Higgs scalar $\F$, its
covariant derivative $D_\mu\F$, and of course the curvature $F_{\mu\nu}$.

Here, we display only the HCS densities in $d=4$, arrived at from the one-step descents of the HCP densities in $D=5$,
each of them descended from the CP densities in $6$ and $8$ dimensions respectively,
\bea
\Omega^{(4,6)}_{\rm HCS}&=&\vep^{\mu\nu\rho\si}\,\mbox{Tr}\ F_{\mu\nu}\,F_{\rho\si}\,\F\label{HCS46}\\
\Omega^{(4,8)}_{\rm HCS}&=&\vep^{\mu\nu\rho\si}\,\mbox{Tr}\bigg[
\F\left(\eta^2\,F_{\mu\nu}F_{\rho\si}+\frac29\,\F^2\,F_{\mu\nu}F_{\rho\si}+\frac19\,F_{\mu\nu}\F^2F_{\rho\si}\right)
\nonumber\\
&&\qquad\qquad\qquad-\frac29
\left(\F D_{\mu}\F D_{\nu}\F-D_{\mu}\F\F D_{\nu}\F+D_{\mu}\F D_{\nu}\F\F\right)F_{\rho\si}\bigg]\,.\label{HCS48}
\eea

Finally, we select the gauge group appropriate for the purpose of transiting from Yang-Mills to gravity.
As stated at the outset by \re{con1}, this group is $SO(D)$ ($D=2n$),
the orthogonal group of the non-Abelian field defining the Chern-Pontryagin (CP) density from which
the Chern-Simons (CS) density is derived, and in the case of Higgs-CS, the gauge group
of the Higgs-CP density from which the HCS is derived.

In the case of CS densities the Dirac matrix representations, $e.g.$ \re{L}, are employed,
such that the spin-connection and the $SO(D)$ YM
connections are identified, $A_\mu^{ab}=\om_\mu^{ab}$. 

The situation is somewhat more involved in the case of Higgs-CS (HCS) densities,
in which case we have HCS densities both in odd deimensions, $e.g.$ \re{HCS36}, \re{HCS38}, \re{HCS58}, \re{HCS710} and \re{CS32},
and in even dimensions $e,g.$ \re{HCS46} and \re{HCS48}. Here, $D$ is even for the HCS in odd dimensions but it
is even when the HCS is in even dimensions. Besides, in this case the multiplicity of the Higgs scalar $\F$
must alo be chosen~\footnote{These choices coincide with those for which monopoles on $\R^D$ are constructed~\cite{Tchrakian:2010ar}.}.
For HCS densities in odd dimensions, $i.e.$ with even $D$, \re{con1} is augmented by the choice of Higgs multiplet,
\be
\label{evenDoddd}
A_{\mu}=-\frac12\,A_{\mu}^{ab}\,\ga_{ab}\ ,\quad {\rm and}\quad\F=2\f^a\,\ga_{a,D+1}\,,
\ee
while for HCS densities in even dimensions, $i.e.$ with $D$ odd,
\be
\label{evendoddD}
A_{\mu}=-\frac12\,A_{\mu}^{ab}\,\Si_{ab}\ ,\quad {\rm and}\quad\F=2\f^a\,\Si_{a,D+1}\,,
\ee
where $\Si^{(\pm)}_{ab}$ are one or other chiral representations of $SO(D)$,
\be
\label{sigma}
\Si^{(\pm)}_{ab}=\frac12\left(\eins\pm\ga_{D+1}\right)\,\ga_{ab}\ ,\quad a=1,2,\dots D\,.
\ee

It is in order to remark that for odd $d$ only the CS densities $\Om_{\rm CS}^{(d)}$, \re{CS32}-\re{CS72},
appear in the HCS densities $\Om_{\rm HCS}^{(d,N)}$,
\re{HCS36}-\re{HCS710}. The CS densities $\hat\Om_{\rm CS}^{(d)}$, \re{CS31}-\re{CS71} do not appear in the latter.

\section{Chern-Simons gravity (CSG) and Higgs-CS gravity (HCSG)}
Both the CSG and the HCSG result in gravitational systems consisting of the superposition of $p$-Einstein-Hilbert densities \re{pEH}
which we present here to be self-comtained and to fix the notation.

In terms of the the spin-connection $\om_M^{ab}$, the covariant derivative of (some frame-vector valued field) $\f^a$ is defined as
\be
D_M\f^a=\pa_M\f^a+\om_M^{ab}\f^b\label{cov}
\ee
and employing further the $vielbein$ field $e_M^a$ there follow the definitions of the
gravitational curvature and torsion
\bea
R_{\mu\nu}^{ab}&=&(D_{[\mu}D_{\nu]})^{ab}=\pa_\mu\om_\nu^{ab}
-\pa_\nu\om_\mu^{ab}+\om_\mu^{ac}\om_\nu^{cb}-\om_\nu^{ac}\om_\mu^{cb}\,\label{Rcurv}\\
C_{\mu\nu}^a&=&D_{[\mu}e_{\nu]}^a=\pa_{\mu}e_{\nu}^a-\pa_{\nu}e_{\mu}^a+\om_\mu^{ac}e_\nu^c-\om_\nu^{ac}e_\mu^c\,,\label{tor}
\eea
with
\be
\label{index1}
\mu=1,2,\dots,d\ ;\quad a=1,2,\dots,d\,.
\ee
To define the $p$-Einsten-Hilbert~\footnote{Aka. Lovelock gravitity.}
($p$-EH) Lagrangians, we split the indices on the Levi-Civita symbols as follows
\[
\vep^{\mu_1\mu_2\dots \mu_{2p}\mu_{2p+1}\dots \mu_d}\ \quad{\rm and}\quad\vep_{a_1a_2\dots a_{2p}a_{2p+1}\dots a_d}
\]
such that in $d$-dimensional spacetime the Lagrangians are
\be
\label{pEH}
{\cal L}^{(p,d)}_{\rm EH}=
\vep^{\mu_1\mu_2\dots \mu_{2p}\mu_{2p+1}\dots \mu_d}\,e_{\mu_{2p+1}}^{a_{2p+1}}e_{\mu_{2p+2}}^{a_{2p+2}}\dots e_{\mu_d}^{a_d} 
\vep_{a_1a_2\dots a_{2p}a_{2p+1}\dots a_d}\,R_{\mu_1\mu_2}^{a_1a_2}R_{\mu_3\mu_4}^{a_3a_4}\dots R_{\mu_{2p-1}\mu_{2p}}^{a_{2p-1}a_{2p}}
\ee
For $p=0$ in $d=2p$ this is a total divergence, while for $d\ge 2p$ it is the cosmological constant.
For $p=1$ it is the usual Einstein-Hilbert (EH) Lagrangian in $d$-dimensions, for $p=2$ it is the usual
Gauss-Bonnet Lagrangian in $d$-dimensions, $etc$.

The definitions of \re{pEH} include the Levi-Civita symbol in both the frame indices and
the coordinate indices. Thus it is appropriate to adopt the definitions \re{CS32}-\re{CS72}
for the CS densities since in that case evaluating the traces will result in some superpositions of (usual) $p$-EH Lagrangians.
In this respect, the choice of \re{HCS36}-\re{HCS710} for the HCS densities is the appropriate one.
Examples of CSG and HCSG systems are given in
the next Subsections, respectively.

\subsection{Chern-Simons gravity (CSG): odd dimensions}
As remarked earlier, the prescription \re{con1} for transiting from YM to gravity in $d$ dimensions,
yield a frame-index $a=1,2,\dots d+1$ which
is defective. This defect is overcome by splitting $a$ as $a=(\al,d+1)=(\al,D)$, with $\al=1,2,\dots, d$,
as will be described in the following Subsections.

In this case the prescription \re{con1}, or the first member of \re{evenDoddd}, is refined as follows,
\bea
A_{\mu}&=&-\frac12\,\om_{\mu}^{\al\bt}\,\ga_{\al\bt}+\ka\,e_{\mu}^{\al}\,\ga_{\al D}\Rightarrow
F_{\mu\nu}=
-\frac12\left(R_{\mu\nu}^{\al\bt}-\ka^2\,e_{[\mu}^{\al}\,e_{\nu]}^{\bt}\right)\ga_{\al\bt}+\ka\,C_{\mu\nu}^{\al}\ga_{\al D}\label{oddd}
\eea
where $C_{\mu\nu}^{\al}=D_{[\mu}e_{\nu]}$ is the torsion. Clearly, $\al$ is now the frame-index with the correct range. The constant
$\ka$ in \re{oddd} has dimensions $L^{-1}$ to compensate for the difference in the dimensions of the connection and the $Vielbein$.

Substituting \re{oddd} in $\Om^{n}_{\rm CS}$, \re{CS32}-\re{CS72}, yields the CSG models in $d=3,5,7$,
\bea
{\cal L}_{\rm CSG}^{(3)}&=&
-\ka\,\vep^{\mu\nu\la}\vep_{abc}\left(R_{\mu\nu}^{ab}-\frac23\,\ka^2e_{\mu}^ae_{\nu}^b\right)e_{\la}^c\label{3csg}\\
{\cal L}_{\rm CSG}^{(5)}&=&
\ka\,\vep^{\mu\nu\rho\si\la}\vep_{abcde}\left(\frac34\,R_{\mu\nu}^{ab}\,R_{\rho\si}^{cd}-\ka^2\,R_{\mu\nu}^{ab}\,e_{\rho}^ce_{\si}^d
+\frac35\,\ka^4e_{\mu}^ae_{\nu}^be_{\rho}^ce_{\si}^d\right)e_{\la}^e\label{5csg}\\
{\cal L}_{\rm CSG}^{(7)}&=&
-\ka\,\vep^{\mu\nu\rho\si\tau\ka\la}\vep_{abcdefg}\bigg(\frac18\,R_{\mu\nu}^{ab}\,R_{\rho\si}^{cd}\,R_{\tau\ka}^{ef}
-\frac14\,\ka^2\,R_{\mu\nu}^{ab}\,R_{\rho\si}^{cd}\,e_{\tau}^ee_{\ka}^f\nonumber\\
&&\qquad\qquad\qquad\qquad+\frac{3}{10}\,\ka^4\,R_{\mu\nu}^{ab}\,e_{\rho}^ce_{\si}^de_{\tau}^ee_{\ka}^f
-\frac17\,\ka^6e_{\mu}^ae_{\nu}^be_{\rho}^ce_{\si}^de_{\tau}^ee_{\ka}^f\bigg)e_{\la}^g\label{7csg}
\eea
Each of these is a linear sum of all the $p$-Einstein-Hilbert (EH) Lagrangians ${\cal L}_{\rm EH}^{(p,d)}$ 
defined in the given dimension $d$ (the $p=0$ member being the cosmological constant term.).

In the notation of Appendix {\bf A},
\bea
{\cal L}_{\rm CSG}^{(3)}&=&-\ka\,\left[\tau_{(1)}{\cal L}_{\rm EH}^{(1,3)}-\tau_{(0)}\ka^2{\cal L}_{\rm EH}^{(0,3)}
\right]\label{csg3x}\\
{\cal L}_{\rm CSG}^{(5)}&=&\ka\,\left[\tau_{(2)}{\cal L}_{\rm EH}^{(2,5)}-\tau_{(1)}\ka^2{\cal L}_{\rm EH}^{(1,5)}
+\tau_{(0)}\ka^4{\cal L}_{\rm EH}^{(0,5)}
\right]\label{csg5x}\\
{\cal L}_{\rm CSG}^{(7)}&=&
-\ka\,\left[\tau_{(3)}{\cal L}_{\rm EH}^{(3,7)}-\tau_{(2)}\ka^2{\cal L}_{\rm EH}^{(2,7)}
+\tau_{(1)}\ka^4{\cal L}_{\rm EH}^{(1,7)}-\tau_{(0)}\ka^6{\cal L}_{\rm EH}^{(0,7)}
\right]\label{csg7x}
\eea
where the dimensionless constants $\tau_{(p)}$ can be read off \re{3csg}, \re{5csg} and \re{7csg}.

\subsection{Higgs--Chern-Simons gravity (HCSG): all dimensions}
The HCS densities, displayed in Subsection {\bf 2.2}, involve both the YM and the Higgs fields.
The passage of the YM-Higgs (YMH) system to gravity is prescribed by \re{evenDoddd} in odd dimensions, and \re{evendoddD} in even.
We will henceforth refer to the case of odd $d=D-1$, namely to the prescription \re{oddd}, since 
for even $d$ the appropriate prescription can be read off \re{oddd}
by formally replacing $(\ga_{\al\bt},\ga_{\al D})$ with $(\Si_{\al\bt},\Si_{\al D})$, the latter defined by \re{sigma}.

As in the previous Subsection, the index $a=(\al,D)$ is split such that $\al$ now is the frame-index, and the refined version of
the second mamber of \re{evenDoddd} we apply is
\bea
2^{-1}\F&=&(\f^{\al}\,\ga_{\al,D+1}+\f\,\ga_{D,D+1})\Rightarrow\nonumber\\
&\Rightarrow& 2^{-1}D_{\mu}\F=(D_{\mu}\f^{\al}-\ka\,e_\mu^\al\,\f)\ga_{\al,D+1}+(\pa_{\mu}\f+\ka\,e_\mu^\al\,\f^\al)\ga_{D,D+1}
\label{higgsevenD2}
\eea
where
\be
\label{gravcov}
D_{\mu}\f^{\al}=\pa_{\mu}\f^{\al}+\om_{\mu}^{\al\bt}\f^{\bt}
\ee
is the gravitational covariant derivative.

We employ \re{oddd} and \re{higgsevenD2} to calculate the traces in the HCS formulas in Subsection {\bf 2.2}. Here, we display only the
pair of HCSG (gravitational) systems arising from HCS densities \re{HCS36}-\re{HCS38} in $d=3$, and the pair arising from
\re{HCS46}-\re{HCS48} in $d=4$.

The pair in $d=3$ is
\bea
{\cal L}_{\rm HCSG}^{(3,6)}&=&\vep^{\la\mu\nu}\vep_{\al\bt\ga}\Bigg\{2\eta^2\ka\,\left(e_{\la}^{\ga}\,R_{\mu\nu}^{\al\bt}
-\frac23\ka^2e_{\mu}^{\al}e_{\nu}^{\bt}e_{\la}^{\ga}\right)\nonumber\\
&&-\bigg[2(R_{\mu\nu}^{\al\bt}-\ka^2\,e_{[\mu}^{\al}e_{\nu]}^{\bt})\left[\f^{\ga}(\pa_{\la}\f+\ka\,e_{\la}^d\f^\del)
-\f (D_{\la}\f^{\ga}-\ka\, e_{\la}^\ga\f)\right]\nonumber\\
&&\qquad\qquad\qquad\qquad\qquad\qquad-4\ka\f^{\al}(D_{\la}\f^{\bt}-\ka\,e_\la^\bt\f)C_{\mu\nu}^{\ga}\bigg]\Bigg\}
\label{HCSG36}
\\
{\cal L}_{\rm HCSG}^{(3,8)}&=&
-3\eta^2{\cal L}_{\rm HGCS}^{(3,3)}-12\vep^{\la\mu\nu}\vep_{\ga\al\bt}\left[\eta^2-(|\f^{\del}|^2+\f^2)\right]\cdot\nonumber\\
&&\qquad\qquad\qquad\cdot\bigg[\left[\f^{\ga}(\pa_{\la}\f+\ka\,e_{\la}^d\f^\del)
-\f (D_{\la}\f^{\ga}-\ka\, e_{\la}^\ga\f)\right]\nonumber\\
&&\qquad\qquad\qquad\qquad\qquad\qquad-4\ka\f^{\al}(D_{\la}\f^{\bt}-\ka\,e_\la^\bt\f)C_{\mu\nu}^{\ga}\bigg]
\label{HCSG38}
\eea
and the pair in $d=4$ is
\bea
{\cal L}_{\rm HCSG}^{(4,6)}&=&
-\vep^{\mu\nu\rho\si}\vep_{\al\bt\ga\del}\,\f\left[R_{\mu\nu}^{\al\bt}R_{\rho\si}^{\ga\del}
-4\ka^2\,e_{\rho}^{\ga}e_{\si}^{\del}R_{\mu\nu}^{\al\bt}+4\ka^4\,\,e_{\mu}^{\al}e_{\nu}^{\bt}e_{\rho}^{\ga}e_{\si}^{\del}\right]
\nonumber\\
&&\qquad+2\ka\vep^{\mu\nu\rho\si}\vep_{\al\bt\ga\del}\left(R_{\mu\nu}^{\al\bt}
-\ka^2\,e_{[\mu}^{\al}\,e_{\nu]}^{\bt}\right)C_{\rho\si}^{\ga}\f^{\del}\label{HCSG46}\\
{\cal L}_{\rm HCSG}^{(4,8)}&=&\left[\eta^2-\frac13(|\f^{\al}|^2+\f^2)\right]{\cal L}_{\rm HGCS}^{(4,6)}-\nonumber\\
&&-\frac23\vep^{\mu\nu\rho\si}\vep_{\al\bt\ga\del}\left[\left(R_{\mu\nu}^{\al\bt}
-\ka^2\,e_{[\mu}^{\al}\,e_{\nu]}^{\bt}\right)\f_{\rho}^{\ga}(\f\f_{\si}^{\del}-2\f^{\del}\f_{\si})
+\frac23C_{\mu\nu}^{\al}\f^{\bt}\f_{\rho}^{\ga}\f_{\si}^{\del}\right]
\label{HCSG48}
\eea
where we have used an abbreviated notation
\[
\f_\mu=\pa_\mu\f\ ,\quad\f_\mu^\al=D_\mu\f^\al\ , \quad\mu=1,2,\dots d\ ,\quad \al=1,2,\dots d\,.
\]

Some concluding remarks
are now in order. We observe the following qualitative properties of the listed HCSG (gravitaional) systems:
\begin{itemize}
\item
In the pair of models in $d=3$, namely \re{HCSG36}-\re{HCSG38}, the leading terms are
the usual Einstein-Hilbert Lagrangian ${\cal L}_{\rm EH}^{(1,3)}$, $viz.$ \re{3csg} or \re{csg3x}.
By contrast, in \re{HCSG46} and \re{HCSG48}, no purely gravitational Lagrangians ${\cal L}_{\rm EH}^{(p,d)}$ appear without the presence
of the frame-vector field $\f^\al$ and the scalar $\f$. This is not surprising, since there exist no CS densities in even dimensions.
\item
The models ${\cal L}_{\rm HGCS}^{(d,N)}$ pertaining to higher values of $N$ in the HCS densities
$\Om_{\rm HCS}^{(d,N)}$ from which they follow, feature Lagrangians with lower $N$, nested inside.
\item
All HCSG models, in odd and even dimensions, feature the torsion term explicitly. This, together with the fact that they
feature the  gravitational covariant derivative \re{gravcov}, means that these models can sustain non-zero torsion.
Whether or not torsion-free solutions may exist, must be checked in each case.
\item
The frame-vector field $\f^\al$ and the scalar $\f$ are relics of the Higgs scalar in the Yang-Mills--Higgs systems giving
rise to the HCSG models.
Thus we would expect that these are gravitational coordinates and not matter fields that might result in hairy solutions.
Accordingly, we would expect that these models support only black hole solutions, and not regular ones.
\end{itemize}

\section{Gravitational CS (GCS)  densities}
Gravitational CS (GCS) densities, as their name suggests, are not gravitational models like the CSG and HCSG models discussed above.
They are the analogues of the non-Abelian (nA) CS densities, that are employed in various applications of nA gauge theories.
Like the latter, the GCS densities are designed to find application in the same way, in gravitational theories.

Together with the CSG models discussed above, GCS densities are derived from the nA CS densities
by applying the prescription \re{con1}, but not
in the refined versions \re{oddd}. As result the frame indices $a=1,2,\dots,D$, $D=d+1$, in the resulting gravitational density
have the wrong range, namely that they range over $\al=1,2,\dots,d$.
This defect is corrected by introducing a (rather aribitary) truncation, which consists of setting
some components of the spin-connection $\om_{\mu}^{ab}=(\om_{\mu}^{\al\bt},\om_{\mu}^{\al,D})$ equal to zero by hand according to
\be
\label{trunc}
\om_{\mu}^{\al,D}=0\quad\Rightarrow\quad R_{\mu\nu}^{\al,D}=0\,.
\ee
The resulting density is expressed exclusively in terms of the components of the (gravitational) connection and curvature
$(\om_\mu^{\al\bt},R_{\mu\nu}^{\al\bt})$, such that now the frame-indices $\al$ transform with the required group $SO(d)$
and not $SO(D)$. This is adopted as the definition of gravitational Chern-Simons (GCS) density.

As in Section {\bf 3}, one has again the choice~\footnote{Recall that previously in the derivation of the CSG models, the choice of
\re{CS32}-\re{CS72} was made since the Levi-Civita symbol
with frame indices, which results from the presence of the chiral matrix $\ga_{D,D+1}$ under the trace,
was required for the description of gravitational systems. Here, we have no such constraint.} of opting for the definitions
\re{CS32}-\re{CS72}, or, \re{CS31}-\re{CS71} for the nA CS densities, which prior to implementing the truncation \re{trunc} are the CS
densities for gauge group $SO(D)$. A further important distinction form the nA case arises here in the gravitational case
when the choice \re{CS31}-\re{CS71} is made for the nA CS densities.
As a result of gamma-matrix identities~\footnote{The identity in question, in $2n$ dimensions is
\[
\ga_{a_1a_2\dots a_nb_1b_2\dots b_{2n}}
=\delta_{a_1a_2\dots a_n}^{b_1b_2\dots b_{2n}}\eins+\vep_{a_1a_2\dots a_nb_1b_2\dots b_{2n}}\ga_{2n+1}
\]
where $\ga_{a_1a_2\dots a_nb_1b_2\dots b_{2n}}$ is the totally antisymmetrised product of $2n$ gamma matrices in $2n$ dimensions, and
$\ga_{2n+1}$ is the chiral matrix.
},
it turns ot that substituting \re{con1}-\re{L} in \re{CS31}-\re{CS71}, these traces vanish itentically in all $4p-3$ dimensions.
As a result, with this choice gravitational CS (GCS) densities can be constructed
{\bf only} in $4p-1$, and {\bf not}, all odd dimensions.
The choice of \re{CS31}-\re{CS71} is for the nA CS densities on the other hand, is not subject to this obstacle and it affords the
definition of CSG densities in all odd dimensions. In this case the resulting CSG density will feature the Levi-Civita symbol
with frame indices, which subject to the truncation \re{trunc}, collapses.

Succinctly stated, CSG densities thus constructed, exist in $4p-1$ dimensionst only. Applying the prescription in $d=3,5,7$,
\be
\label{prescr1}
A_{\mu}=-\frac12\,\om_{\mu}^{ab}\,\ga_{ab}\ \,,\quad a=1,2,\dots,d+1\,,
\ee
namely by eveluating the traces in \re{CS31}-\re{CS71} and then implementing the truncation \re{trunc},
\bea
\hat\Om_{\rm GCS}^{{(3)}}&=&-\frac{1}{2\cdot2!}\,\vep^{\la\mu\nu}\del_{\al\bt}^{\bar\al\bar\bt}\
\om_{\la}^{\al\bt}\left[R_{\mu\nu}^{\bar\al\bar\bt}-\frac23\left(\om_{\mu}\om_{\nu}\right)^{\bar\al\bar\bt}\right]\label{GCS31}\\
\hat\Om_{\rm GCS}^{{(5)}}&=&0\label{GCS51}\\
\hat\Om_{\rm GCS}^{{(7)}}&=&\frac{1}{2\cdot6!}\,\vep^{\la\mu\nu\rho\si\tau\ka}\hat\del_{\al\bt\ga\del}^{\bar\al\bar\bt\bar\ga\bar\del}
\,\om_{\la}^{\al\bt}\bigg[R_{\mu\nu}^{\ga\del}R_{\rho\si}^{\bar\al\bar\bt}R_{\tau\ka}^{\bar\ga\bar\del}
-\frac45R_{\mu\nu}^{\ga\del}R_{\rho\si}^{\bar\al\bar\bt}\left(\om_{\tau}\om_{\ka}\right)^{\bar\ga\bar\del}-\frac25
R_{\mu\nu}^{\ga\del}\left(\om_{\rho}\om_{\si}\right)^{\bar\al\bar\bt}R_{\tau\ka}^{\bar\ga\bar\del}\nonumber\\
&&\qquad\qquad\qquad\quad
+\frac45R_{\mu\nu}^{\ga\del}\left(\om_{\rho}\om_{\si}\right)^{\bar\al\bar\bt}\left(\om_{\tau}\om_{\ka}\right)^{\bar\ga\bar\del}
-\frac{8}{35}\left(\om_{\mu}\om_{\nu}\right)^{\ga\del}\left(\om_{\rho}\om_{\si}\right)^{\bar\al\bar\bt}
\left(\om_{\tau}\om_{\ka}\right)^{\bar\ga\bar\del}\bigg]\label{GCS71}
\eea
$etc.$, where the symbol $\hat\del_{\al\bt\ga\del}^{\bar\al\bar\bt\bar\ga\bar\del}$ in \re{GCS71} is
\be
\label{symbol}
\hat\del_{\al\bt\ga\del}^{\bar\al\bar\bt\bar\ga\bar\del}=\frac19\,\del_{\al\bt\ga\del}^{\bar\al\bar\bt\bar\ga\bar\del}
+\frac14\,\del_{\al\bt}^{\ga\del}\del_{\bar\al\bar\bt}^{\bar\ga\bar\del}\,.
\ee

\subsection{Gravitational Higgs-CS (GHCS) densities}
The main purpose of constructing GHCS densities would be to supply GCS densities in both odd and even dimensions,
possibly including in $4p-3$ dimensions which were absent in the Higgs free case above. Thus here too we employ the Higgs-CS (HCS)
densities presented in Section {\bf 2.2}.

In Section {\bf 3}, where CS and HCS gravities were constructed, it turned out that in all odd dimensions
the leading term in the HCS gravity (HCSG) was the CS gravity (CSG). In that case the choice of CS densities \re{CS32}-\re{CS72}
displaying the chiral matrix $\ga_{D+1}=(\ga_{d+2})$, was made with the aim of generating a gravitational model, which coincided with
\re{HCS36}-\re{HCS710}, the defining of the HCS densities in odd dimensions.

If we invoke the same criterion here as in the construction of HCS gravities (HCSG),
namely that the leading terms in the GHCS densities in $4p-1$
dimensions be the CSG densities, $e.g.$ $\hat\Om_{\rm CS}^{(d)}$ given by \re{CS31} and \re{CS71} in $d=3,7$, then this is achieved by
deforming \re{HCS36}-\re{HCS38} and \re{HCS710}, by removing the chiral matrix under
the trace, $by\ hand$.

The corresponding consideration in $d=4p-3$ dimensions fails to yield a nontrivial result.
We know the GCS density $\hat\Om_{\rm CS}^{(5)}$ vanishes, $cf.$ \re{CS51}.

To illustrate these points, consider the examples of proposed (deformed) HCS densities in $d=3,5$
\bea
\hat\Omega^{(3,6)}_{\rm HCS}&=&-2\eta^2\hat\Omega_{\rm CS}^{(3)}
-\vep^{\mu\nu\la}\mbox{Tr}\,D_{\la}\F\left(F_{\mu\nu}\,\F+F_{\mu\nu} \F\right)\,.\label{hatHCS36}\\
\hat\Omega^{(5,8)}_{\rm HCS}&=&2\eta^2\hat\Om_{\rm CS}^{(5)}+\vep^{\mu\nu\rho\si\la}\,\mbox{Tr}\,
D_{\la}\F\left(\F F_{\mu\nu}F_{\rho\si}+F_{\mu\nu}\F F_{\rho\si}+F_{\mu\nu}F_{\rho\si}\F\right)\label{hatHCS58}
\eea
Applying the prescription
\be
\label{prescr2}
A_{\mu}=
-\frac12\,\om_{\mu}^{ab}\,\ga_{ab}\ ,\quad {\rm and}\quad\F=2\f^a\,\ga_{a,d+2}\,,\quad a=1,2,\dots,d+1\,,
\ee
and then implementing the
truncation
\re{trunc}, we find the two GHCS densities
\bea
\hat\Om_{\rm GHCS}^{(3,6)}&=&-2\eta^2\hat\Om_{\rm GCS}^{{(3)}}+4\vep^{\la\mu\nu}\,\f^\al R_{\mu\nu}^{\al\bt}D_\la\f^\bt\label{GHCS36}\\
\hat\Om_{\rm GHCS}^{(5,8)}&=&0+0\label{GHCS58}
\eea
As we see from \re{GHCS58}, the gravitational HCS (GHCS) densities vanish in $4p-3$ dimensions, just as the 
gravitational CS (GCS) densities, for the same technical reason. (Not deforming the HCS density by removing the chiral matrix
from under the trace does not change the situation. In that case the Levi-Civita symbol in the frame indices in $D=d+1$ dimensions
appear, which vanish when the trauncation \re{trunc} is implemented.)

Concerning the construction of GHCS densities in even dimensions, we employ the HCS densities
$\Om_{\rm GHCS}^{(4,6)}$ and $\Om_{\rm GHCS}^{(4,8)}$ given by \re{HCS46}-\re{HCS48} in $d=4$. The prescription applied here
is also \re{prescr2}, but with $\ga_{ab}$ formally replaced by $\Si_{ab}$, $cf.$ \re{sigma}, followed by the truncation \re{trunc}.
The result is
\bea
\hat\Om_{\rm GHCS}^{(4,6)}&=&
-\frac14\,\vep^{\mu\nu\rho\si}\,R_{\mu\nu}^{\al\bt}\,R_{\rho\si}^{\al\bt}\,\f\label{GHCS46}\\
\hat\Om_{\rm GHCS}^{(4,8)}&=&
-\vep^{\mu\nu\rho\si}\,
R_{\mu\nu}^{\al\bt}\left\{\left[\frac18\left(1-\frac{1}{3}|\f^a|^2\right)R_{\rho\si}^{\al\bt}
+\frac13\,\f_{\rho\si}^{\al\bt}\right]\f+\frac43\f^{\al}D_{\rho}\f^{\bt}\,\pa_{\si}\f\right\}\label{GHCS48}
\eea
where the abbreviated notation
\[
\f_{\mu\nu}^{\al\bt}=D_{[\mu}\f^\al D_{\nu]}\f^\bt\ , \quad{\rm and}\quad \f=\f^5\,.
\]

In even dimensions, there are no exclusions like in odd dimensions, and GHCS densities like \re{GHCS46} and \re{GHCS48}
exist in all even dimensions.

Some concluding remarks are in order here.
\begin{itemize}
\item
In odd dimensions, gravitational CS (GCS) densities and gravitational HCS (GHCS) densities are defined only in $4p-1$ and not
in $4p-3$ dimensions.
\item
In $4p-1$ dimensions, the leading term in the GHCS density is the GCS density.
\item
Gravitational HCS (GHCS) densities can be defined in all even dimensions where no GCS densities exist.
\item
As in the case of HCS gravities (HCSG), the frame-vector field $\f^\al$ and the scalar $\f$ are gravitational degrees of freedom.
\item
If employed as CS densities to modify a gravitational model, the GHCS densities would be applied to HCSG gravitational models
decsribed in Section {\bf 3}, which are decsribed by the same gravitational fields. This is because the fields $\f^a=(\f^\al,\f)$
are gravitational degrees of freedom and their dynamics is given naturally by the HCSG models in the given dimension.
\end{itemize}

\section{Summary}
An illustraive presentation of Chern-Simons gravities in all dimensions, and gravitational Chern-Simons densities in all even and
in $4p-1$ odd dimensions is given. These ``Chern-Simons densities'' are one-step descendents of
(Higgs--)Chern-Pontryagin densities defined in all dimensions, and which result
from the dimensional decsent of a Chern-Pontryagin density in some (higher) even dimension. A distiction is made between CS
gravitational systems and the CS densities, and each is presented separately followed by comments in its own Section. 

\bigskip
\noindent
{\bf Acknowledgements}: My deepest gratitude to Eugen Radu for his unstinting support in preparing this report. My thanks
to Friedrich Hehl for having introduced me to the Einstein-Cartan formulation, and to
Ruben Manvelyan for helpful extended discussions. Thanks to Jorge Zanelli for helpful correspondence.

\begin{small}

\end{small}


\begin{thebibliography}{99}
\bibitem{Witten:1988hc}
  E.~Witten,
  Nucl.\ Phys.\ B {\bf 311} (1988) 46.
  doi:10.1016/0550-3213(88)90143-5
\bibitem{Chamseddine:1989nu}
  A.~H.~Chamseddine,
  Phys.\ Lett.\ B {\bf 233} (1989) 291.
  doi:10.1016/0370-2693(89)91312-9
\bibitem{Chamseddine:1990gk}
  A.~H.~Chamseddine,
  Nucl.\ Phys.\ B {\bf 346} (1990) 213.
  doi:10.1016/0550-3213(90)90245-9
\bibitem{Deser:1982vy}
  S.~Deser, R.~Jackiw and S.~Templeton,
  Phys.\ Rev.\ Lett.\  {\bf 48} (1982) 975.
  doi:10.1103/PhysRevLett.48.975
\bibitem{Araneda:2016iiy}
  R.~Araneda, R.~Aros, O.~Miskovic and R.~Olea,
  Phys.\ Rev.\ D {\bf 93} (2016) no.8,  084022
  doi:10.1103/PhysRevD.93.084022
  [arXiv:1602.07975 [hep-th]].
\bibitem{Deser:1981wh}
  S.~Deser, R.~Jackiw and S.~Templeton,
  Annals Phys.\  {\bf 140} (1982) 372
   [Annals Phys.\  {\bf 281} (2000) 409]
   Erratum: [Annals Phys.\  {\bf 185} (1988) 406].
  doi:10.1006/aphy.2000.6013, 10.1016/0003-4916(82)90164-6
\bibitem{Mardones:1990qc}
  A.~Mardones and J.~Zanelli,
  Class.\ Quant.\ Grav.\  {\bf 8} (1991) 1545.
  doi:10.1088/0264-9381/8/8/018
\bibitem{Zanelli:2012px}
  J.~Zanelli,
  Class.\ Quant.\ Grav.\  {\bf 29} (2012) 133001
  doi:10.1088/0264-9381/29/13/133001
  [arXiv:1208.3353 [hep-th]].
\bibitem{MacDowell:1977jt}
  S.~W.~MacDowell and F.~Mansouri,
  Phys.\ Rev.\ Lett.\  {\bf 38} (1977) 739
   Erratum: [Phys.\ Rev.\ Lett.\  {\bf 38} (1977) 1376].
  doi:10.1103/PhysRevLett.38.1376, 10.1103/PhysRevLett.38.739
\bibitem{Jackiw:2003pm}
  R.~Jackiw and S.~Y.~Pi,
  Phys.\ Rev.\ D {\bf 68} (2003) 104012
  doi:10.1103/PhysRevD.68.104012
  [gr-qc/0308071].
\bibitem{Jackiw:1985}
see for example, R.~Jackiw, "Chern-Simons terms and cocycles in physics and mathematics",
in E.S. Fradkin $Festschrift$, Adam Hilger, Bristol (1985).
\bibitem{Tchrakian:2010ar}
  T.~Tchrakian,
  J.\ Phys.\ A {\bf 44} (2011) 343001
  doi:10.1088/1751-8113/44/34/343001
  [arXiv:1009.3790 [hep-th]].
\bibitem{Radu:2011zy}
  E.~Radu and T.~Tchrakian,
  doi:10.1142/9789814440349-0020
  arXiv:1101.5068 [hep-th].
\bibitem{Tchrakian:2015pka}
  D.~H.~Tchrakian,
  J.\ Phys.\ A {\bf 48} (2015) no.37,  375401
  doi:10.1088/1751-8113/48/37/375401
  [arXiv:1505.05344 [hep-th]].
\bibitem{Forgacs:1979zs}
  P.~Forgacs and N.~S.~Manton,
  Commun.\ Math.\ Phys.\  {\bf 72} (1980) 15.
  doi:10.1007/BF01200108
\bibitem{Schwarz:1981mb}
  A.~S.~Schwarz and Y.~S.~Tyupkin,
  Nucl.\ Phys.\ B {\bf 187} (1981) 321.
  doi:10.1016/0550-3213(81)90277-7
\bibitem{Kapetanakis:1992hf}
  D.~Kapetanakis and G.~Zoupanos,
  Phys.\ Rept.\  {\bf 219} (1992) 4.
  doi:10.1016/0370-1573(92)90101-5
\bibitem{Sherry:1984ky} 
  T.~N.~Sherry and D.~H.~Tchrakian,
  Phys.\ Lett.\  {\bf 147B}, 121 (1984).
  doi:10.1016/0370-2693(84)90605-1
\bibitem{OBrien:1988whk}
  G.~M.~O'Brien and D.~H.~Tchrakian,
  Mod.\ Phys.\ Lett.\ A {\bf 4} (1989) 1389.
  doi:10.1142/S0217732389001581
\bibitem{Tchrakian:1984gq}
  D.~H.~Tchrakian,
  Phys.\ Lett.\  {\bf 150B} (1985) 360.
  doi:10.1016/0370-2693(85)90994-3
\bibitem{Navarro-Lerida:2013pua}
  F.~Navarro-Lerida, E.~Radu and D.~H.~Tchrakian,
  Int.\ J.\ Mod.\ Phys.\ A {\bf 29} (2014) no.26,  1450149
  doi:10.1142/S0217751X14501498
  [arXiv:1311.3950 [hep-th]].
\bibitem{Navarro-Lerida:2014rwa}
  F.~Navarro-Lerida and D.~H.~Tchrakian,
  Int.\ J.\ Mod.\ Phys.\ A {\bf 30} (2015) no.15,  1550079
  doi:10.1142/S0217751X15500797
  [arXiv:1412.4654 [hep-th]].
\bibitem{Szabo:2014zua}
  R.~J.~Szabo and O.~Valdivia,
  JHEP {\bf 1406} (2014) 144
  doi:10.1007/JHEP06(2014)144
  [arXiv:1404.4319 [hep-th]].
\bibitem{Romanov:1978ur}
  V.~N.~Romanov, I.~V.~Frolov and A.~S.~Schwarz,
  Teor.\ Mat.\ Fiz.\  {\bf 37} (1978) 305.
  doi:10.1007/BF01018584
\bibitem{Obukhov:1995eq}
  Y.~N.~Obukhov and F.~W.~Hehl,
  Acta Phys.\ Polon.\ B {\bf 27} (1996) 2685
  [gr-qc/9602014].

\end{thebibliography}
\end{document}